\documentclass[12pt]{article}

\usepackage{amsmath,amssymb,amsfonts,graphics,graphicx,amscd,amsfonts,epsfig,color}
\usepackage{array, booktabs,bm,cite}
\usepackage{subfig}
\usepackage{here}
\newcommand {\beq}{\begin{equation}}
\newcommand {\eeq}{\end{equation}}
\newcommand {\beqa}{\begin{eqnarray}}
\newcommand {\eeqa}{\end{eqnarray}}

\renewcommand{\theequation}{\thesection.\arabic{equation}}

\begin{document}
\setlength{\oddsidemargin}{0cm}
\setlength{\baselineskip}{7mm}

\begin{titlepage}
\renewcommand{\thefootnote}{\fnsymbol{footnote}}
\begin{normalsize}
\begin{flushright}
\begin{tabular}{l}
April 2018
\end{tabular}
\end{flushright}
  \end{normalsize}

~~\\

\vspace*{0cm}
    \begin{Large}
       \begin{center}
         {Renormalization on the fuzzy sphere}
       \end{center}
    \end{Large}
\vspace{1cm}

\begin{center}
           Kohta H{\sc atakeyama}$^{1,2)}$\footnote
            {
e-mail address : 
hatakeyama.kohta.15@shizuoka.ac.jp},
           Asato T{\sc suchiya}$^{1,2)}$\footnote
            {
e-mail address : 
tsuchiya.asato@shizuoka.ac.jp}
           {\sc and}
           Kazushi Y{\sc amashiro}$^{1)}$\footnote
           {
e-mail address : yamashiro.kazushi.17@shizuoka.ac.jp}\\
      \vspace{1cm}

 $^{1)}$ {\it Department of Physics, Shizuoka University}\\
                {\it 836 Ohya, Suruga-ku, Shizuoka 422-8529, Japan}\\
         \vspace{0.3cm}     
        $^{2)}$ {\it Graduate School of Science and Technology, Shizuoka University}\\
               {\it 3-5-1 Johoku, Naka-ku, Hamamatsu 432-8011, Japan}

\end{center}

\hspace{5cm}

\begin{abstract}
\noindent
We study renormalization on the fuzzy sphere. 
We numerically simulate a scalar field theory on it, which is described by a Hermitian matrix model. 
We show that 
correlation functions defined by using the Berezin symbol are made
independent of the matrix size, which is viewed as a UV cutoff, by tuning a parameter of the theory.
We also find that the theories on the phase boundary are universal.
They behave as a conformal field theory at short distances, while
they show an effect of UV/IR mixing at long distances.
\end{abstract}
\vfill
\end{titlepage}
\vfil\eject

\setcounter{footnote}{0}

\section{Introduction}
\setcounter{equation}{0}
\renewcommand{\thefootnote}{\arabic{footnote}} 
A lot of attention has been paid to field theories on noncommutative spaces, 
mainly because they have a deep connection to string theory or quantum gravity (for a review, see \cite{Douglas:2001ba}.).
One of the most peculiar phenomena in field theories on noncommutative spaces
is the so-called UV/IR mixing \cite{Minwalla:1999px}.
This is known to be an obstacle to perturbative renormalization.

In \cite{Chu:2001xi,Steinacker:2016nsc}, 
the UV/IR mixing in a scalar field theory on the fuzzy sphere\footnote{The theory has been studied by Monte Carlo simulation in \cite{Martin:2004un,Panero:2006bx,Panero:2006cs,
GarciaFlores:2009hf,Das:2007gm,Hatakeyama:2017fao}. For related analytic studies of the model,
see\cite{Kawamoto:2015qla,Vaidya:2003ew,OConnor:2007ibg,Nair:2011ux,
Polychronakos:2013nca,Tekel:2013vz,Saemann:2014pca,Tekel:2014bta,Tekel:2015zga}.}, which is
realized by a matrix model, was examined 
perturbatively:
the one-loop self-energy 
differs from that in the ordinary theory
on a sphere by finite and non-local terms 
even in the commutative limit.
This effect is sometimes called the UV/IR anomaly.

It is important to elucidate the problem of renormalization to construct consistent quantum field theories on noncommutative spaces.
It was shown in \cite{Hatakeyama:2017fao} by Monte Carlo study
that by tuning the mass parameter 
the 2-point and 4-point correlation functions in the disordered phase
of the above theory
are made independent of
the matrix size up to a wave function renormalization, where the matrix size
is interpreted as a UV cutoff\footnote{A similar analysis for a scalar field theory on the noncommutative torus was performed in \cite{Bietenholz:2004xs,Mejia-Diaz:2014lza}}.
This strongly suggests that the theory is nonperturbatively renormalizable in the 
disordered phase.

In this paper, we perform further study of the scalar field theory
on the fuzzy sphere by Monte Carlo simulation.
We define the correlation functions by using the Berezin symbol \cite{Berezin:1974du}
as in \cite{Hatakeyama:2017fao}.
First, we show that the 2-point and 4-point correlation functions are made independent
of the matrix size by tuning the coupling constant. 
Thus, we verify a conjecture in \cite{Hatakeyama:2017fao} that the theory is universal up to a parameter fine-tuning.
Next, 
we identify the phase boundary by measuring the susceptibility that is an order parameter for the $Z_2$ symmetry
and calculate the 2-point and 4-point correlation functions on the boundary. 
We find that the correlation functions at different points on the boundary agree 
so that the theories on the boundary are universal as in ordinary field theories. 
Furthermore, we observe that the 2-point correlation functions behave as those in a conformal field theory (CFT) at short distances but deviate from it at long distances. 
It is nontrivial that the behavior of the CFT is seen because field theories on noncommutative spaces are non-local ones.

This paper is organized as follows.
In section 2, we introduce the scalar field theory on the fuzzy sphere and review
its connection to the theory on a sphere.
In section 3, we study renormalization in the disordered phase.
In section 4, we identify the phase boundary 
and calculate the 2-point and 4-point 
correlation functions on the boundary.
Section 5 is devoted to the conclusion and discussion.
In the appendix, we review the Bloch coherent state and the Berezin symbol.

\section{Scalar field theory on the fuzzy sphere}
\setcounter{equation}{0}
Throughout this paper, we examine the following matrix model:
\begin{equation}
S=\frac{1}{N}\mbox{Tr}\left(-\frac{1}{2}[L_i,\Phi]^2+\frac{\mu^2}{2}\Phi^2
+\frac{\lambda}{4}\Phi^4\right) \ ,
\label{action}
\end{equation}
where $\Phi$ is an $N\times N$ Hermitian matrix, and
$L_i$ ($i=1,2,3$) are the generators of the $SU(2)$ algebra with the spin-$(N-1)/2$ representation, which obey the commutation relation
\begin{equation}
[L_i,L_j]=i\epsilon_{ijk}L_k \ .
\end{equation}
The theory possesses $Z_2$ symmetry: $\Phi \rightarrow -\Phi$.
The path-integral measure is given by $d\Phi e^{-S}$, where
\begin{equation}
d\Phi=\prod_{i=1}^N d\Phi_{ii} \prod_{1\leq j < k\leq N}
d\mbox{Re}\Phi_{jk} d\mbox{Im}\Phi_{jk} \ .
\end{equation}

The theory (\ref{action}) reduces to the following continuum theory
on a sphere with the radius $R$
at the tree level in the $N\rightarrow \infty$ limit, which corresponds
to the so-called commutative limit:
\begin{equation}
S_c=\frac{R^2}{4\pi}\int d\Omega \left(-\frac{1}{2R^2}({\cal L}_i\phi)^2
+\frac{m^2}{2}\phi^2+\frac{g}{4}\phi^4\right) \ ,
\label{continuum action}
\end{equation}
where $d\Omega$ is the invariant measure on the sphere and 
${\cal L}_i$ ($i=1,2,3$) are the orbital angular momentum operators. 
The correspondence of the parameters in (\ref{action}) and (\ref{continuum action})
is given by
\begin{align}
\mu^2 & = R^2 m^2 \ , \nonumber\\
\lambda & = R^2 g \ .
\label{m^2 and lambda}
\end{align}
We review the above tree-level correspondence in the appendix.
It was shown in \cite{Chu:2001xi,Steinacker:2016nsc} 
that there exist finite differences between
 (\ref{action}) and (\ref{continuum action}) in the perturbative expansion,
which are known as the UV/IR anomaly.

To define correlation functions,
we introduce the Berezin symbol \cite{Berezin:1974du} that is constructed from
the Bloch coherent state \cite{Gazeau}.
We parametrize the sphere in terms of the standard polar coordinates 
$\Omega=(\theta,\varphi)$.
The Bloch coherent state $|\Omega \rangle$
is localized around the point $(\theta,\varphi)$ with the width $R/\sqrt{N}$.
The Berezin symbol for an $N\times N$ matrix $A$ is given by $\langle\Omega|A|\Omega\rangle$.
The Berezin symbol $\langle \Omega | \Phi | \Omega \rangle$ is identified
with the field $\phi(\Omega)$ 
in the correspondence at the tree level between 
(\ref{action}) and (\ref{continuum action}).
The Bloch coherent state and the Berezin symbol are reviewed in the appendix.

\section{Correlation functions}
\setcounter{equation}{0}
\subsection{Definition of correlation functions}
By denoting the Berezin symbol briefly as 
\begin{equation}
\varphi(\Omega) = \langle\Omega | \Phi | \Omega\rangle \ ,
\end{equation}
we define the $n$-point correlation function in the theory (\ref{action}) as
\begin{equation}
\left\langle \varphi(\Omega_1)\varphi(\Omega_2)\cdots\varphi(\Omega_n)
\right\rangle
=\frac{\int d\Phi \ \varphi(\Omega_1)\varphi(\Omega_2)\cdots\varphi(\Omega_n) 
\ e^{-S}}{\int d\Phi \  e^{-S}} \ .
\label{n-point function}
\end{equation}
The correlation function (\ref{n-point function}) is a counterpart of 
$\langle \phi(\Omega_1)\phi(\Omega_2)\cdots\phi(\Omega_n)\rangle$ in
the theory (\ref{continuum action}).

We assume that the matrix $\Phi$ in (\ref{action}) is renormalized as
\begin{equation}
\Phi=\sqrt{Z}\Phi_r \ ,
\end{equation}
where $\Phi_r$ is the renormalized matrix.
Then, we define the renormalized Berezin symbol $\varphi_r(\Omega)$ by
\begin{equation}
\varphi(\Omega)=\sqrt{Z}\varphi_r(\Omega) \ ,
\end{equation}
and the renormalized $n$-point correlation function 
$\left\langle \varphi_r(\Omega_1)\varphi_r(\Omega_2)\cdots\varphi_r(\Omega_n)
\right\rangle$ by
\begin{equation}
\left\langle \varphi(\Omega_1)\varphi(\Omega_2)\cdots\varphi(\Omega_n)
\right\rangle
=Z^{\frac{n}{2}}
\left\langle \varphi_r(\Omega_1)\varphi_r(\Omega_2)\cdots\varphi_r(\Omega_n)
\right\rangle \ .
\end{equation}

In the following, we calculate the following correlation functions:
\begin{align}
&\mbox{1-point function:} \; \left\langle \varphi(\Omega_1) \right\rangle  \ ,
\nonumber\\
&\mbox{2-point function:} \; \left\langle \varphi(\Omega_p)\varphi(\Omega_q)
\right\rangle \; (1\leq p < q\leq 4) \ , \nonumber\\
&\mbox{4-point function:} \; \left\langle \varphi(\Omega_1)\varphi(\Omega_2)
\varphi(\Omega_3)\varphi(\Omega_4) \right\rangle \ .
\label{correlation functions}
\end{align}
We verify that the 1-point functions vanish in the parameter region 
that we examine in this section.
This implies that we work in the disordered phase.
Thus, the 2-point correlation functions are themselves the connected ones, while
the connected 4-point correlation functions are given by
\begin{align}
\left\langle \varphi(\Omega_{1})  \varphi(\Omega_{2}) 
\varphi(\Omega_{3})  \varphi(\Omega_{4}) \right\rangle_c 
=&
\left\langle \varphi (\Omega_{1})  \varphi (\Omega_{2}) \varphi
(\Omega_{3})  \varphi (\Omega_{4}) \right\rangle 
- \left\langle \varphi(\Omega_{1})  \varphi(\Omega_{2}) \right\rangle 
\left\langle \varphi(\Omega_{3})  \varphi(\Omega_{4}) \right\rangle \nonumber \\
&- \left\langle \varphi(\Omega_{1})  \varphi(\Omega_{3}) \right\rangle 
\left\langle \varphi(\Omega_{2})  \varphi(\Omega_{4}) \right\rangle 
- \left\langle \varphi(\Omega_{1})  \varphi(\Omega_{4}) \right\rangle 
\left\langle \varphi(\Omega_{2})  \varphi(\Omega_{3}) \right\rangle \ ,
\end{align}
where $c$ stands for the connected part.
The renormalized correlation functions are defined as
\begin{align}
\left\langle \varphi(\Omega_{1}) \right\rangle 
&= \sqrt{Z} \left\langle \varphi_r(\Omega_{1}) \right\rangle \ , \\
\left\langle \varphi(\Omega_{p})  \varphi(\Omega_{q}) \right\rangle 
&= Z \left\langle \varphi_r (\Omega_{p})  \varphi_r (\Omega_{q}) \right\rangle \ , \\
\left\langle \varphi(\Omega_{1})  \varphi(\Omega_{2}) 
\varphi(\Omega_{3})  \varphi(\Omega_{4}) \right\rangle_c 
&= Z^2 \left\langle \varphi_r (\Omega_{1})  \varphi_r (\Omega_{2})  
\varphi_r (\Omega_{3})  \varphi_r (\Omega_{4}) \right\rangle_c  \ .
\end{align}
\begin{figure}[t]
\centering
\includegraphics[width=5cm]{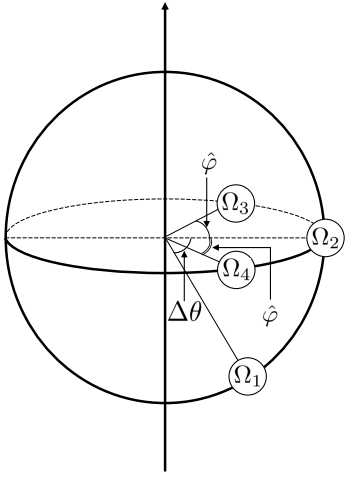}
\caption{Four points on the sphere chosen for the correlation functions. }
\label{sphere}
\end{figure}
We pick up four points $\Omega_p=(\theta_p,\varphi_p)$ ($p=1,2,3,4$) on the sphere as follows 
(see Fig. \ref{sphere}):
\begin{align}
\Omega_1 &= \left(\frac{\pi}{2} + \Delta \theta, \ 0\right)  \ , \nonumber\\
\Omega_2 &= \left(\frac{\pi}{2} , \ 0\right)  \ , \nonumber\\
\Omega_3 &= \left(\frac{\pi}{2} , \ \hat{\varphi}\right) \ , \nonumber\\
\Omega_4 &= \left(\frac{\pi}{2} , \ -\hat{\varphi}\right) \ ,
\label{angles on the sphere}
\end{align}
where $\hat{\varphi}=\pi/12$ and $\Delta \theta=0.1m$ with $m$ taken from $1$ to $15$.

We apply the hybrid Monte Carlo method to our simulation of the theory.

\subsection{Tuning $\mu^2$}
In this subsection, we renormalize the theory by tuning $\mu^2$.
We fix $\lambda$ at $1.0$.

\begin{figure}[t]
\centering
\includegraphics[width=11cm]{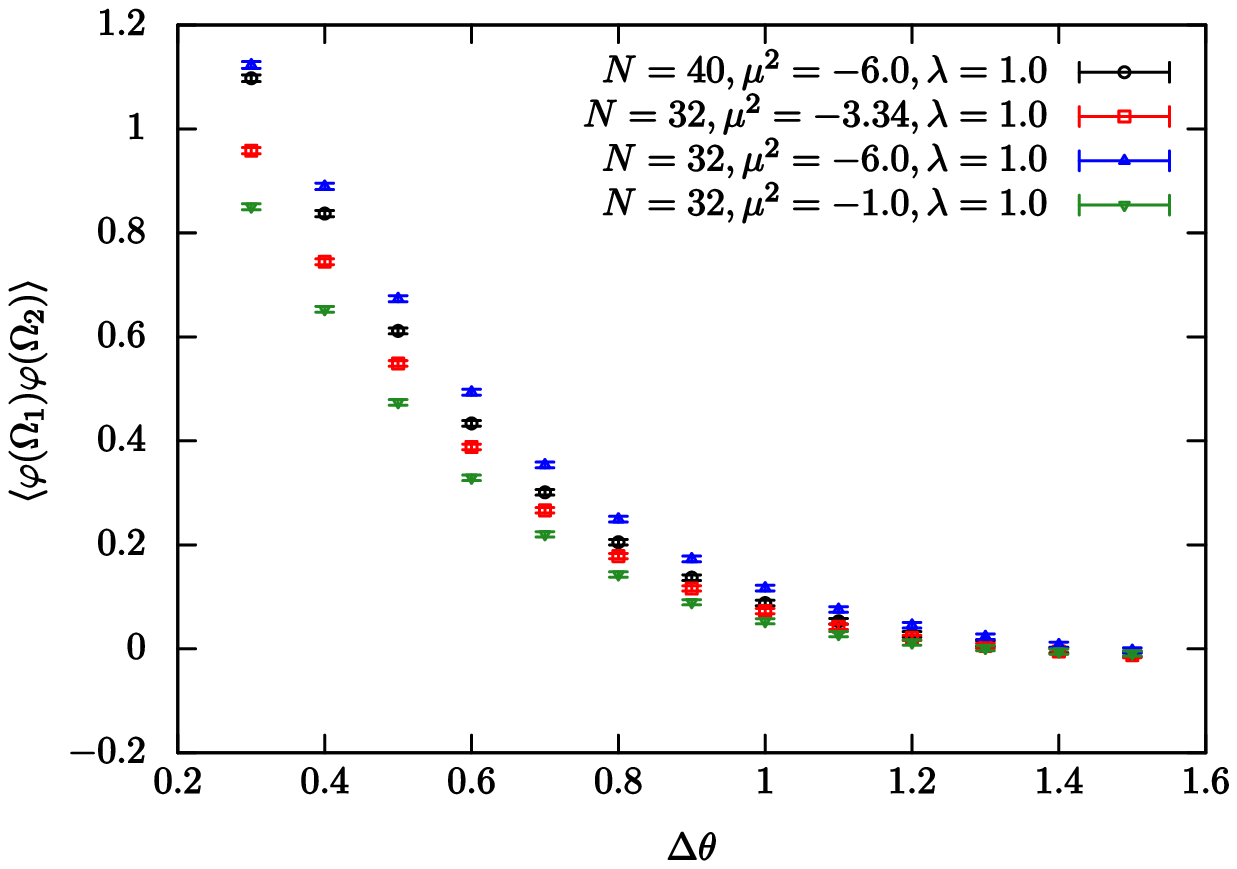}
\caption{$\left\langle\varphi(\Omega_1)\varphi(\Omega_2)\right\rangle$
at $\lambda=1.0$ is
plotted against $\Delta\theta$. Circles represent the data for $N=40$ and $\mu^2=-6.0$, 
while squares, triangles, and inverted triangles represent the data for
$N=32$ and $\mu^2=-3.34, -6.0, -1.0$, respectively.}
\label{2pt_32_lambda_fixed}
\end{figure}

\begin{figure}
\centering
\includegraphics[width=11cm]{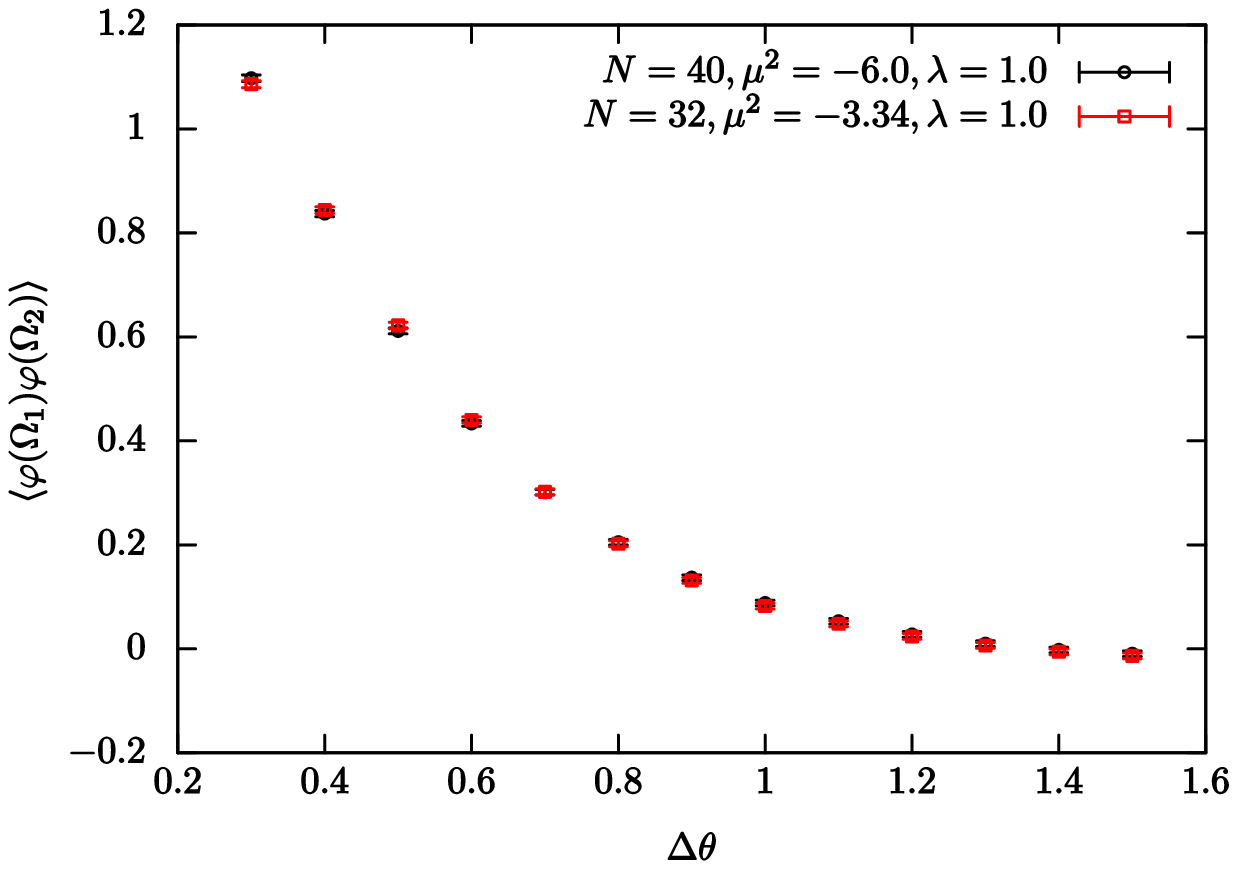}
\caption{$\left\langle\varphi(\Omega_1)\varphi(\Omega_2)\right\rangle$
at $N=40$, $\mu^2=-6.0$, and $\lambda=1.0$ is 
plotted against $\Delta\theta$ (circles).
$\zeta_{32\rightarrow 40}
\left\langle\varphi(\Omega_1)\varphi(\Omega_2)\right\rangle$
with $\zeta_{32\rightarrow 40}=1.263(8)$
at $N=32$, $\mu^2=-3.34$, and $\lambda=1.0$ 
is also plotted against $\Delta\theta$ (squares).}
\label{2pt_32_lambda_fixed_2}
\end{figure}

\begin{figure}
\centering
\includegraphics[width=11cm]{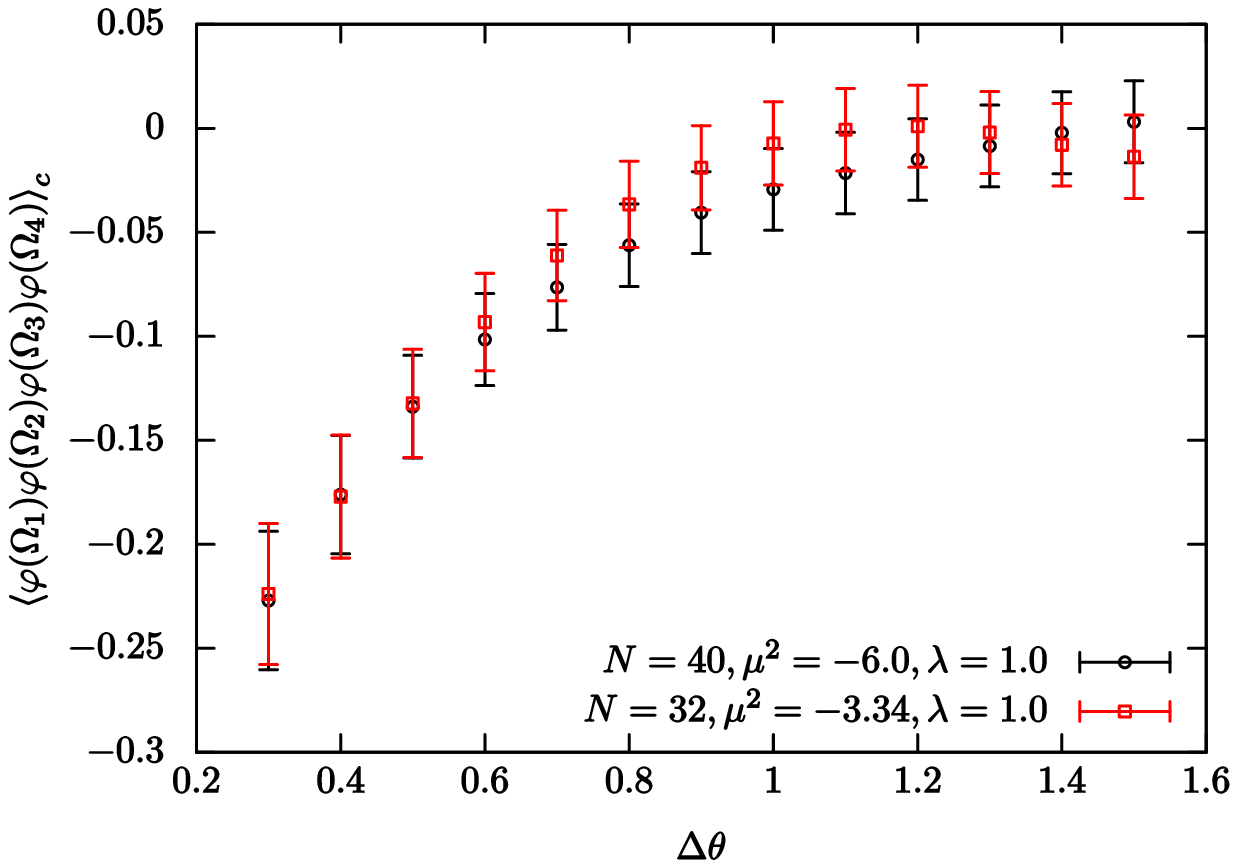}
\caption{$\left\langle \varphi(\Omega_{1})  \varphi(\Omega_{2}) 
\varphi(\Omega_{3})  \varphi(\Omega_{4}) \right\rangle_c$ 
at $N=40$, $\mu^2=-6.0$, and $\lambda=1.0$ is 
plotted against $\Delta\theta$ (circles).
$\zeta_{32\rightarrow 40}^2
\left\langle \varphi(\Omega_{1})  \varphi(\Omega_{2}) 
\varphi(\Omega_{3})  \varphi(\Omega_{4}) \right\rangle_c$
with $\zeta_{32\rightarrow 40}^2=1.595$
at $N=32$, $\mu^2=-3.34$, and $\lambda=1.0$ 
is also plotted against $\Delta\theta$ (squares).}
\label{4pt_32_lambda_fixed}
\end{figure}
First, we simulate at $N=40$ and $\mu^2=-6.0$.
Then, we simulate at $N=32$ for various values of $\mu^2$.
In Fig.\ref{2pt_32_lambda_fixed}, we plot
\begin{equation} 
\left\langle\varphi(\Omega_1)\varphi(\Omega_2)\right\rangle
=Z \left\langle \varphi_r(\Omega_1)\varphi_r(\Omega_2) \right\rangle
\end{equation}
against $\Delta\theta$ at $N=40$ and $\mu^2=-6.0$
and at $N=32$ and typical values of $\mu^2$, $-6.0, -3.34, -1.0$.
We find that the data for $N=32$ and $\mu^2=-3.34$ agree with the ones
for $N=40$ and $\mu^2=-6.0$ if the former are multiplied by a constant
and that this is not the case for the data for $N=32$ and $\mu^2=-6.0, -1.0$.
We determined the above constant as
$\zeta_{32\rightarrow 40}=\frac{Z(40)}{Z(32)}=1.263(8)$ 
by using the least-squares method.
In Fig.\ref{2pt_32_lambda_fixed_2}, 
we plot 
$\left\langle\varphi(\Omega_1)\varphi(\Omega_2)\right\rangle$ 
at $N=40$ and $\mu^2=-6.0$ and 
$\zeta_{32\rightarrow 40}
\left\langle\varphi(\Omega_1)\varphi(\Omega_2)\right\rangle$ at
$N=32$ and $\mu^2=-3.34$ against $\Delta\theta$.
We indeed see that the data for $N=32$ agree nicely with the ones for $N=40$. 
This implies that the renormalized 2-point functions at $N=32$ and $N=40$ agree.

Furthermore,  
in Fig.\ref{4pt_32_lambda_fixed}, we plot
$\left\langle \varphi(\Omega_{1})  \varphi(\Omega_{2}) 
\varphi(\Omega_{3})  \varphi(\Omega_{4}) \right\rangle_c$ at
$N=40$ and $\mu^2=-6.0$ and $\zeta_{32\rightarrow 40}^2
\left\langle \varphi(\Omega_{1})  \varphi(\Omega_{2}) 
\varphi(\Omega_{3})  \varphi(\Omega_{4}) \right\rangle_c$ 
at $N=32$ and $\mu^2=-3.34$ against $\Delta\theta$.
We again see a nice agreement between the data for $N=32$ and the ones for
$N=40$, which means that the renormalized connected 4-point functions at $N=32$
agree with those at $N=40$.
We do not see the above agreement of the correlation functions for $m=1, 2$ in \eqref{angles on the sphere}. We consider this to be attributed to the UV cutoff.

The above results strongly suggest that the correlation functions are
made 
independent of $N$ up to a wave function renormalization by tuning $\mu^2$
and that the theory is nonperturbatively renormalizable in the ordinary sense.


\subsection{Tuning $\lambda$}
In this subsection, we renormalize the theory by tuning $\lambda$.
We fix $\mu^2$ at $-6.0$.

\begin{figure}[t]
\centering
\includegraphics[width=11cm]{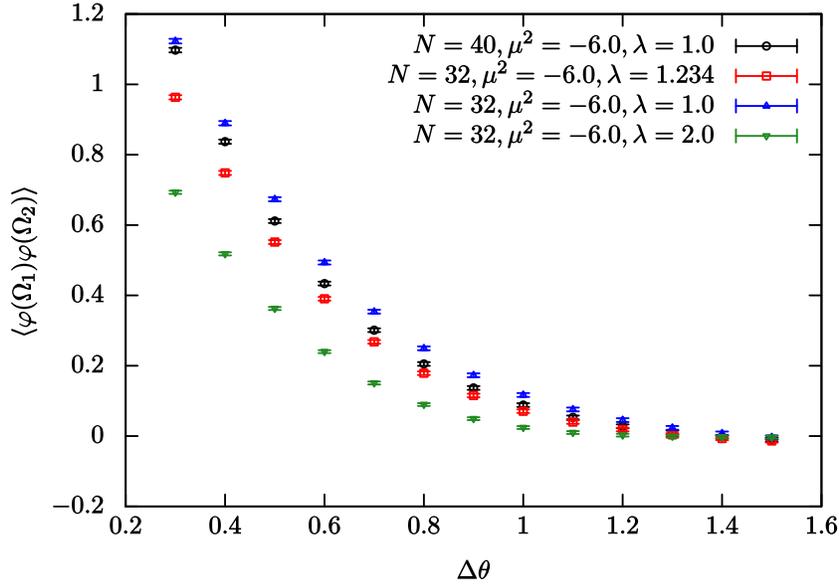}
\caption{$\left\langle\varphi(\Omega_1)\varphi(\Omega_2)\right\rangle$
at $\mu^2=-6.0$ is
plotted against $\Delta\theta$. Circles represent the data for $N=40$ and $\lambda=1.0$, while squares, triangles, and inverted triangles represent the data for
$N=32$ and $\lambda=1.234, 1.0, 2.0$ , respectively. }
\label{2pt_32_mu2_fixed}
\end{figure}

\begin{figure}
\centering
\includegraphics[width=11cm]{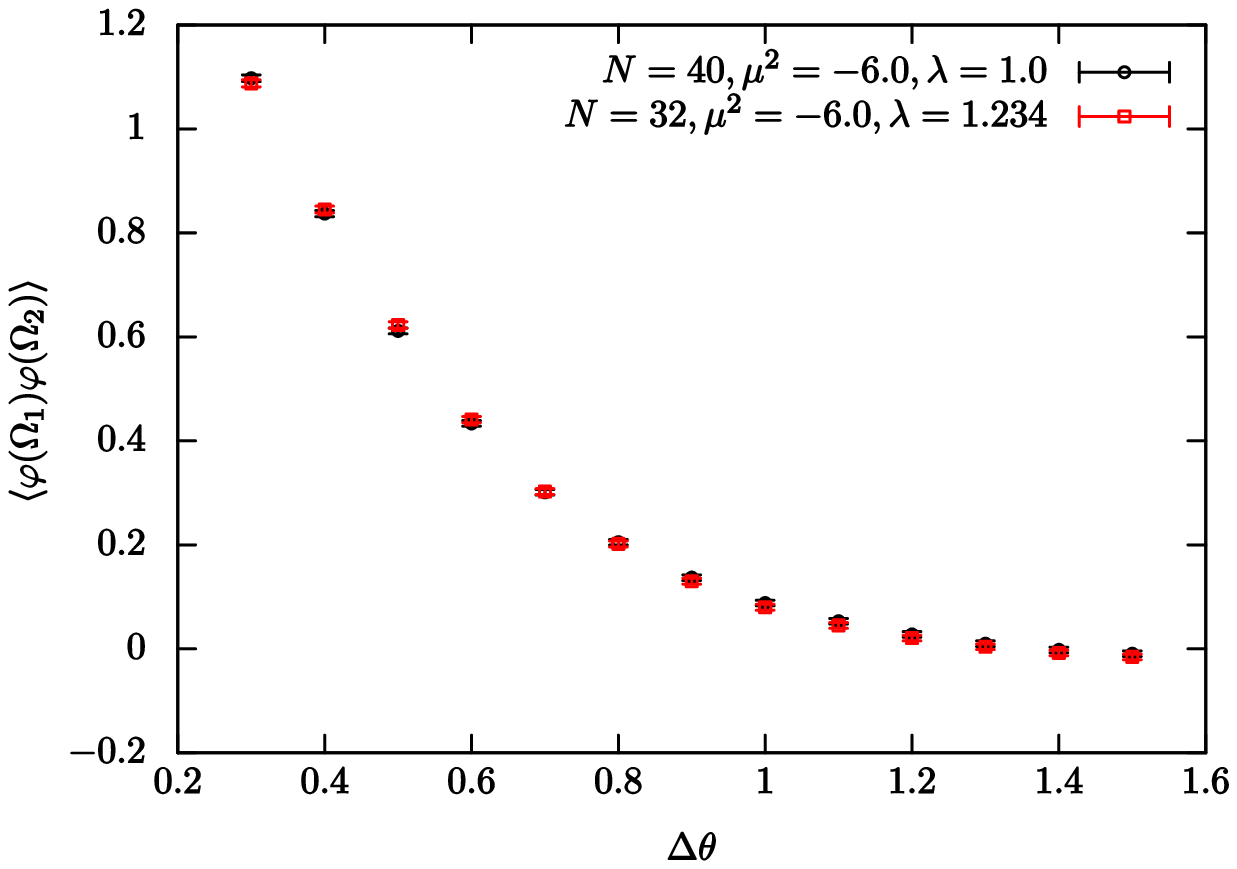}
\caption{$\left\langle\varphi(\Omega_1)\varphi(\Omega_2)\right\rangle$
at $N=40$, $\mu^2=-6.0$, and $\lambda=1.0$ is 
plotted against $\Delta\theta$ (circles).
${\zeta'}_{32\rightarrow 40}
\left\langle\varphi(\Omega_1)\varphi(\Omega_2)\right\rangle$
with ${\zeta'}_{32\rightarrow 40}=1.129(8)$
at $N=32$, $\mu^2=-6.0$, and $\lambda=1.234$ 
is also plotted against $\Delta\theta$ (squares).}
\label{2pt_32_mu2_fixed_2}
\end{figure}

\begin{figure}
\centering
\includegraphics[width=11cm]{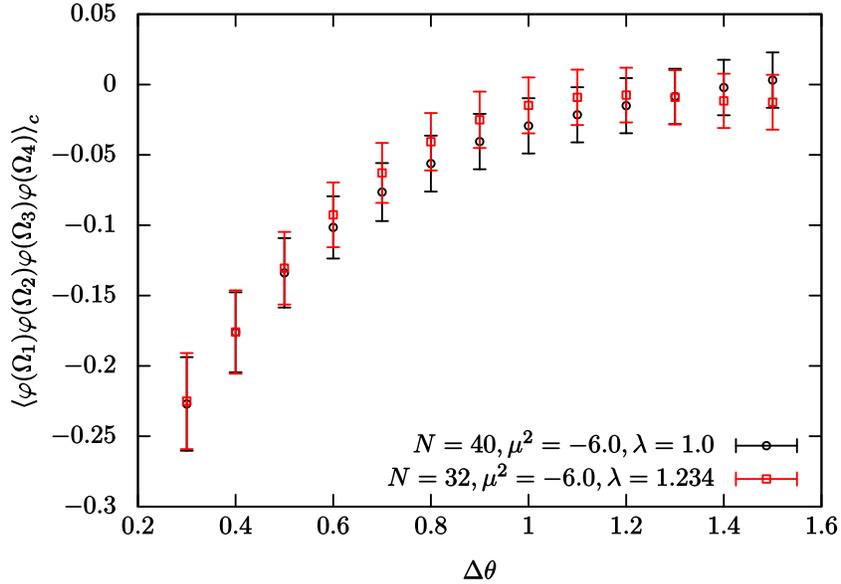}
\caption{$\left\langle \varphi(\Omega_{1})  \varphi(\Omega_{2}) 
\varphi(\Omega_{3})  \varphi(\Omega_{4}) \right\rangle_c$ 
at $N=40$, $\mu^2=-6.0$, and $\lambda=1.0$ is 
plotted against $\Delta\theta$, where circles represent the data.
${\zeta'}_{32\rightarrow 40}^2
\left\langle \varphi(\Omega_{1})  \varphi(\Omega_{2}) 
\varphi(\Omega_{3})  \varphi(\Omega_{4}) \right\rangle_c$
with ${\zeta'}_{32\rightarrow 40}^2=1.275$
at $N=32$, $\mu^2=-6.0$, and $\lambda=1.234$ 
is also plotted against $\Delta\theta$, where squares represent the data.}
\label{4pt_32_mu2_fixed}
\end{figure}
We simulate at $N=32$ 
for various values of $\lambda$.
In Fig.\ref{2pt_32_mu2_fixed}, we plot
\begin{equation} 
\left\langle\varphi(\Omega_1)\varphi(\Omega_2)\right\rangle
=Z \left\langle \varphi_r(\Omega_1)\varphi_r(\Omega_2) \right\rangle
\end{equation}
against $\Delta\theta$ at $N=40$ and $\lambda=1.0$
and at $N=32$ and typical values of $\lambda$, $1.0, 1.234, 2.0$.
We find that the data for $N=32$ and $\lambda=1.234$ agree with the ones
for $N=40$ and $\lambda=1.0$ if the former are multiplied by a constant ${\zeta'}_{32\rightarrow 40}=\frac{Z(40)}{Z(32)}=1.129(8)$
and that this is not the case for
the data for $N=32$ and $\lambda=1.0,2.0$.
In Fig.\ref{2pt_32_mu2_fixed_2}, 
we plot 
$\left\langle\varphi(\Omega_1)\varphi(\Omega_2)\right\rangle$ 
at $N=40$ and $\lambda=1.0$ and 
${\zeta'}_{32\rightarrow 40}
\left\langle\varphi(\Omega_1)\varphi(\Omega_2)\right\rangle$ at
$N=32$ and $\lambda=1.234$ against $\Delta\theta$.
As in the previous section, we see that the data for $N=32$ agree nicely with the ones for
$N=40$. This implies that the renormalized 2-point functions at $N=32$ and $N=40$ agree.

Furthermore, 
in Fig.\ref{4pt_32_mu2_fixed}, we plot
$\left\langle \varphi(\Omega_{1})  \varphi(\Omega_{2}) 
\varphi(\Omega_{3})  \varphi(\Omega_{4}) \right\rangle_c$ at
$N=40$ and $\lambda=1.0$ and ${\zeta'}_{32\rightarrow 40}^2
\left\langle \varphi(\Omega_{1})  \varphi(\Omega_{2}) 
\varphi(\Omega_{3})  \varphi(\Omega_{4}) \right\rangle_c$ 
at $N=32$ and $\lambda=1.234$ against $\Delta\theta$.
We  again see a nice agreement between the data for $N=32$ and the ones for
$N=40$, which means that the renormalized connected 4-point functions at $N=32$
agree with those at $N=40$.

The above results
strongly suggest that the theory is also nonperturbatively renormalized by tuning $\lambda$ in the sense that the renormalized correlation functions are independent of $N$.

The results in the previous and present sections imply that the theory is renormalized by tuning a parameter; namely, it is universal up to a parameter fine-tuning.

\section{Critical behavior of correlation functions}
\setcounter{equation}{0}
In this section, we examine the 2-point and 4-point correlation functions 
on the phase boundary. We fix $N$ at 24 in this section.

We introduce a stereographic projection defined by
\begin{equation}
z = R\tan\frac{\theta}{2} e^{i\varphi} \ ,
\end{equation}
which maps a sphere with the radius $R$ to the complex plane.
Here we fix $R$ at $1$ without loss of generality.
We calculate the 2-point correlation function
\begin{equation}
\langle \varphi(z_m) \varphi(1) \rangle  
\label{2-point function}
\end{equation}
and the connected 4-point correlation function
\begin{equation}
\langle \varphi(z_m) \varphi(1) 
\varphi(e^{i\frac{\pi}{3}})\varphi(e^{i\frac{5\pi}{3}})\rangle_c \ ,
\label{4-point function}
\end{equation}
where 
\begin{equation}
z_m = \tan \left[\frac{1}{2} \left(\frac{\pi}{2} + 0.1m \right) \right]
\end{equation}
with $m$ taken from$1$ to $15$. See Fig. \ref{sphere} with $\hat{\varphi}=\pi/3$.

The renormalized 2-point correlation function
$\langle \varphi_r(z_m) \varphi_r(1) \rangle$
and the renormalized connected 4-point correlation function
$\langle \varphi_r(z_m) \varphi_r(1) 
\varphi_r(e^{i\frac{\pi}{3}})\varphi_r(e^{i\frac{5\pi}{3}})\rangle_c$ are defined by
\begin{align}
\langle \varphi(z_m) \varphi(1) \rangle
&=Z \langle \varphi_r(z_m) \varphi_r(1) \rangle \ , \\
\langle \varphi(z_m) \varphi(1) 
\varphi(e^{i\frac{\pi}{3}})\varphi(e^{i\frac{5\pi}{3}})\rangle_c
&=Z^2 \langle \varphi_r(z_m) \varphi_r(1) 
\varphi_r(e^{i\frac{\pi}{3}})\varphi_r(e^{i\frac{5\pi}{3}})\rangle_c \ .
\end{align}

\begin{figure}[t]
\centering
\includegraphics[width=11cm]{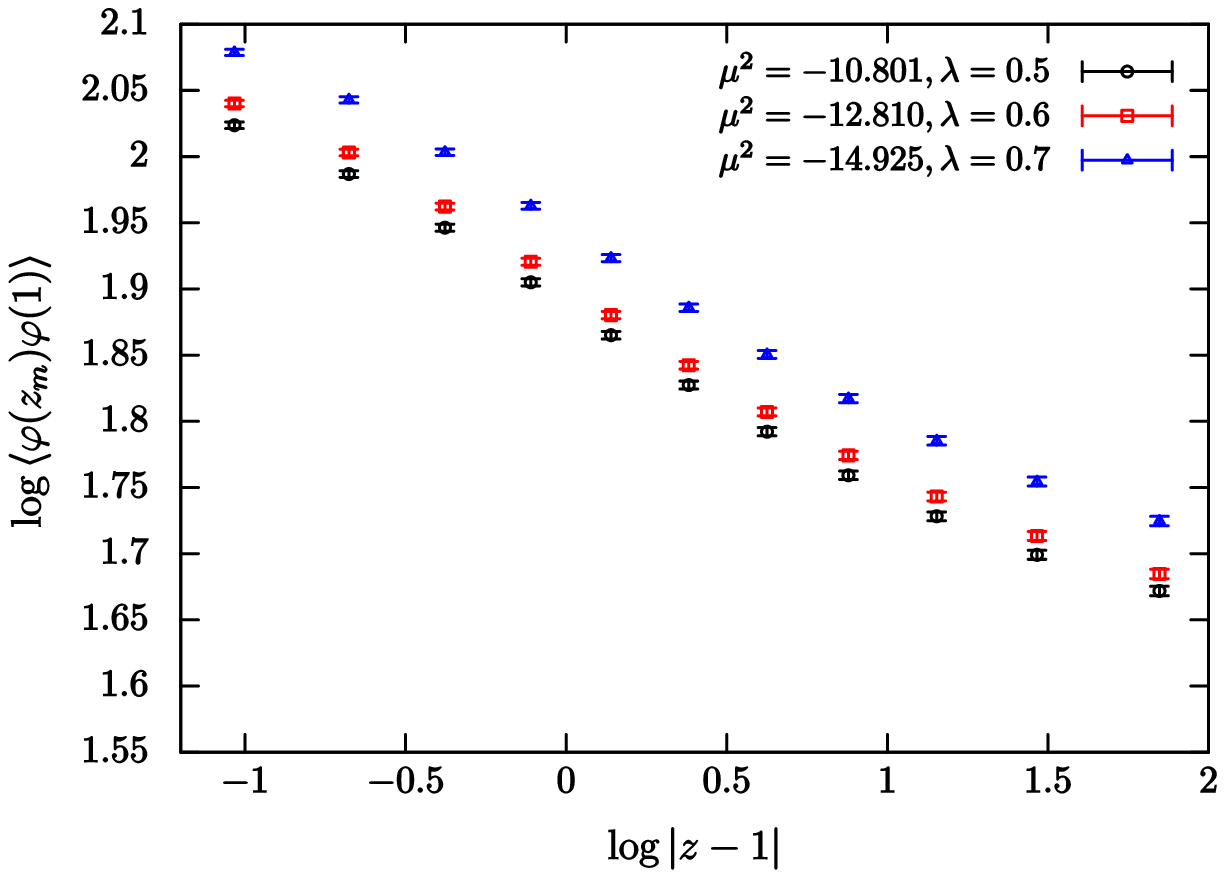}
\caption{$\log \langle \varphi(z_m) \varphi(1) \rangle$ at $N=24$ is plotted against $\log |z-1|$. The data for $(\mu^2, \lambda)=(-10.801, 0.5), (-12.810, 0.6), (-14.925, 0.7)$ are represented by the circles, the squares, and the triangles, respectively.}
\label{2pt_before_renormalization_oncritical}
\end{figure}

\begin{figure}
\centering
\includegraphics[width=11cm]{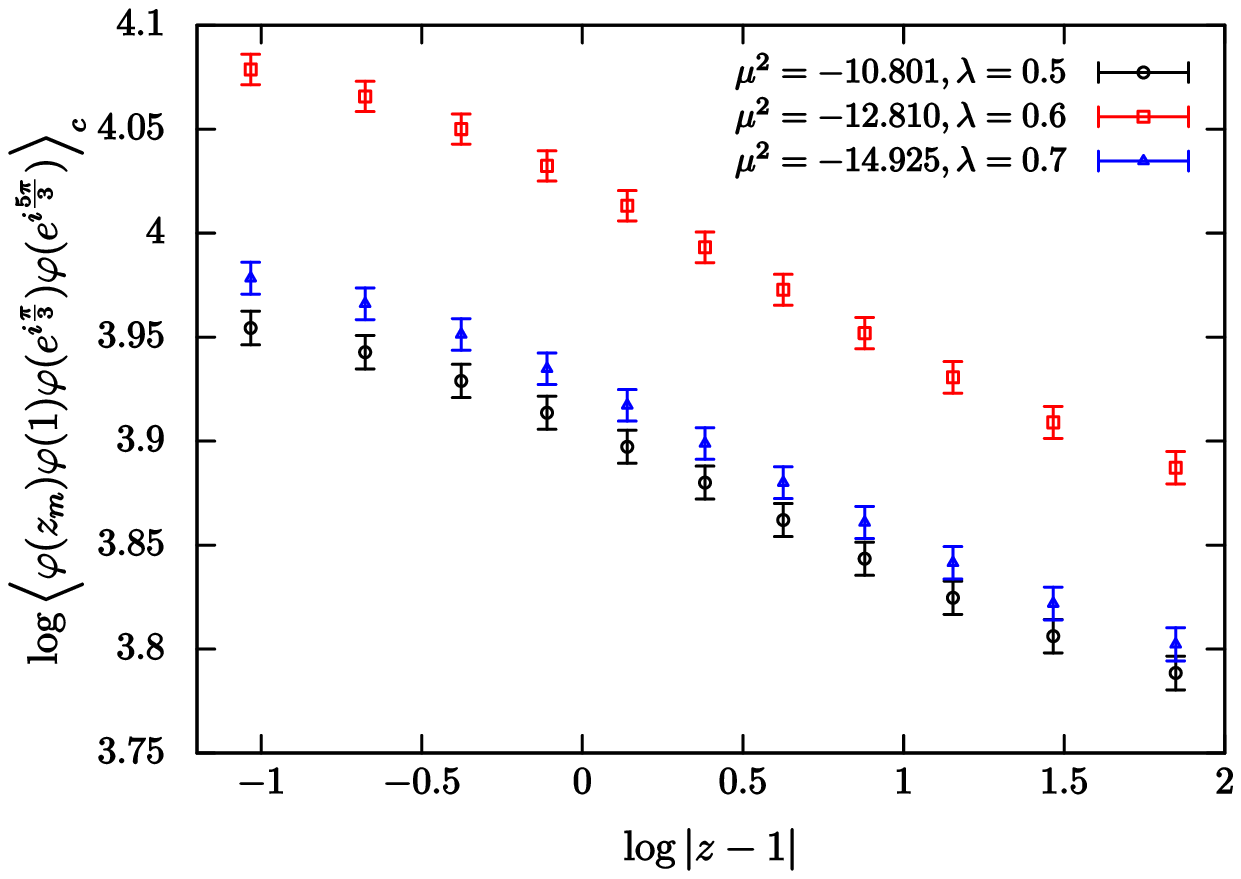}
\caption{$\log \langle \varphi(z_m) \varphi(1) \varphi_r(e^{i\frac{\pi}{3}})\varphi_r(e^{i\frac{5\pi}{3}}) \rangle_c$ at $N=24$ is plotted against $\log |z-1|$. The data for $(\mu^2, \lambda)=(-10.801, 0.5), (-12.810, 0.6), (-14.925, 0.7)$ are represented by the circles, the squares, and the triangles, respectively.}
\label{4pt_before_renormalization_oncritical}
\end{figure}
Here, in order to see a connection to a CFT, we use a log-log plot.
We plot $\log \langle \varphi(z_m) \varphi(1) \rangle$
and $\log \langle \varphi(z_m) \varphi(1) 
\varphi(e^{i\frac{\pi}{3}})\varphi(e^{i\frac{5\pi}{3}})\rangle_c$ against
$\log |z-1|$ for
$(\mu^2, \lambda)=(-10.801, 0.5), (-12.810, 0.6)$, $(-14.925, 0.7)$
in Figs.\ref{2pt_before_renormalization_oncritical} and \ref{4pt_before_renormalization_oncritical}, respectively.
\begin{figure}
\centering
\includegraphics[width=12cm]{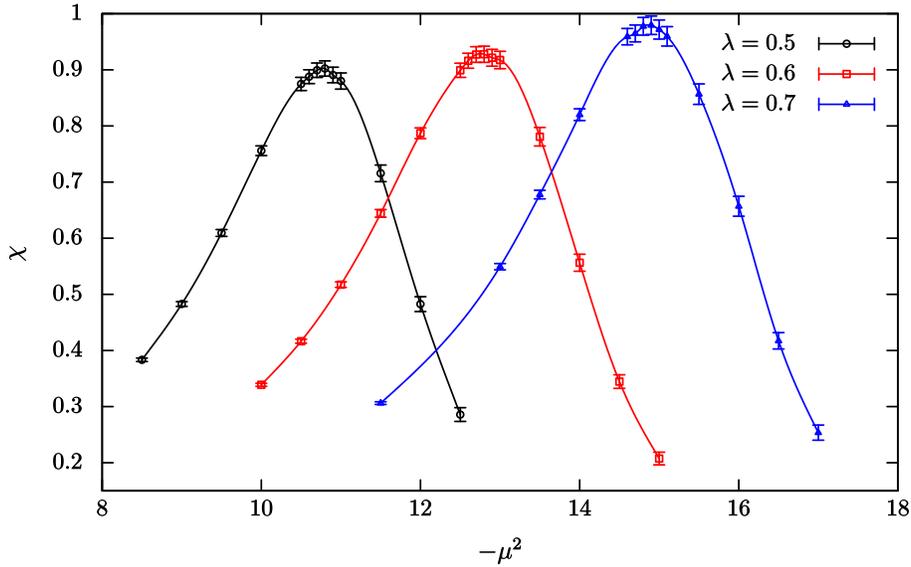}
\caption{The susceptibility $\chi$ at $N=24$ is plotted against $-\mu^2$. The data for $\lambda=0.5, 0.6, 0.7$ are represented by the circles, the squares, and the triangles, respectively. The peaks of $\chi$ for $\lambda=0.5,0.6,0.7$ exist around $\mu^2=-10.8, -12.8, -14.8$, respectively.}
\label{chi}
\end{figure}
\begin{figure}[t]
\centering
\includegraphics[width=11cm]{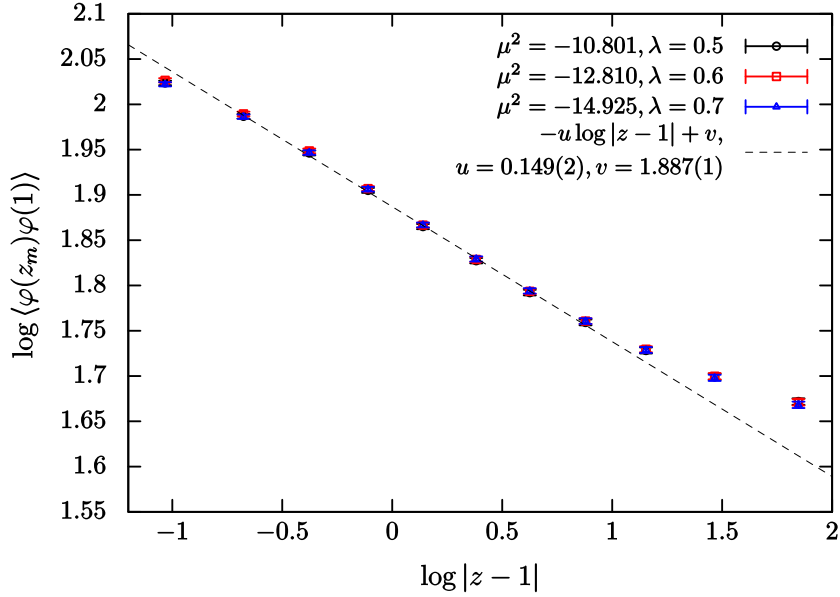}
\caption{$\log \langle \varphi(z_m) \varphi(1) \rangle$ at $N=24$ is plotted against $\log |z-1|$. The data for $(\mu^2, \lambda)=(-10.801, 0.5)$ are the same as in Fig. \ref{2pt_before_renormalization_oncritical}. The data for $(\mu^2, \lambda)=(-12.810, 0.6), (-14.925, 0.7)$ are simultaneously shifted by $\alpha_{0.6 \to 0.5}=-0.015(1)$ and $\alpha_{0.7 \to 0.5}=-0.056(1)$, respectively, in the vertical direction. The data for $(\mu^2, \lambda)=(-10.801, 0.5), (-12.810, 0.6), (-14.925, 0.7)$ are represented by the circles, the squares, and the triangles, respectively. The dashed line is a fit of seven data points (from the second point to the eighth point) of $\log \langle \varphi(z_m) \varphi(1) \rangle$ at $(\mu^2, \lambda)=(-10.801, 0.5)$ to $-u\log |z-1|+v$ with $u=0.149(2)$ and $v=1.887(1)$.}
\label{2pt_after_renormalization_oncritical}
\end{figure}

\begin{figure}[t]
\centering
\includegraphics[width=11cm]{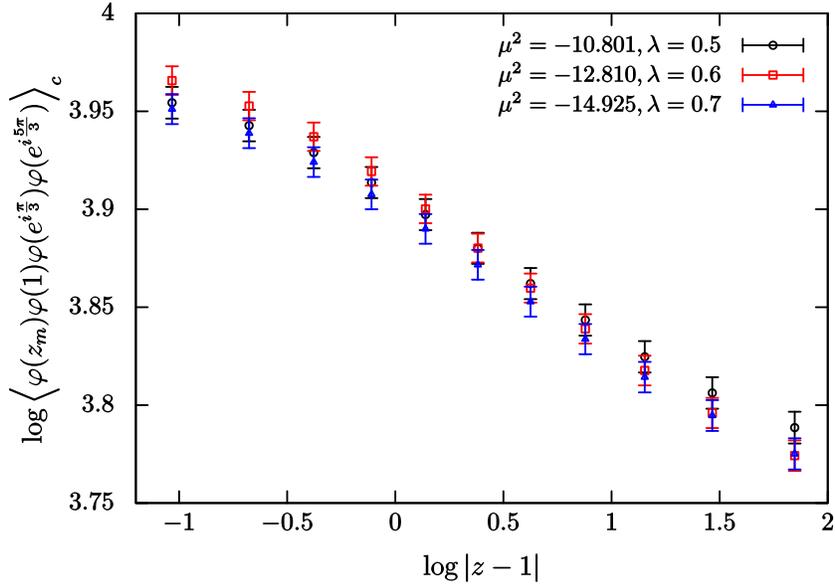}
\caption{$\log \langle \varphi(z_m) \varphi(1) \varphi(e^{i\frac{\pi}{3}})\varphi(e^{i\frac{5\pi}{3}})\rangle_c$ at $N=24$ is plotted against  $\log |z-1|$. 
The data for $(\mu^2, \lambda)=(-10.801, 0.5)$ are the same as in Fig. \ref{4pt_before_renormalization_oncritical}, while the data for $(\mu^2, \lambda)=(-12.810, 0.6), (-14.925, 0.7)$ are simultaneously shifted by $2\alpha_{0.6 \to 0.5}$ and $2\alpha_{0.7 \to 0.5}$, respectively, in the vertical direction. The data for $(\mu^2, \lambda)=(-10.801, 0.5)$, $(-12.810, 0.6), (-14.925, 0.7)$ are represented by the circles, the squares, and the triangles, respectively.}
\label{4pt_after_renormalization_oncritical}
\end{figure}

We also define the susceptibility $\chi$ that is an order parameter for the $Z_2$ symmetry by
\begin{equation}
\chi = \left\langle \left(\frac{1}{N}\mbox{Tr}\Phi \right)^2\right\rangle
-\left\langle \frac{1}{N} \left|\mbox{Tr}\Phi \right| \right\rangle^2 \ .
\end{equation}
In Fig.\ref{chi}, we plot $\chi$ against $-\mu^2$ for each value of $\lambda$, $0.5,0.6,0.7$. 
The critical values of $-\mu^2$, $-\mu^2_c$, that give the peaks of $\chi$ correspond to the phase transition points where symmetry breaking of the $Z_2$ symmetry occurs: the $Z_2$ symmetry is broken for $-\mu^2>-\mu^2_c$, while it is unbroken for $-\mu^2<-\mu^2_c$.
We find that peaks of $\chi$ for $\lambda=0.5,0.6,0.7$ exist around
$\mu^2=-10.8, -12.8, -14.8$, respectively.
We tune the values of $\mu^2$ around the above
values such that the 2-point and 4-point correlation functions for different $\lambda$ agree
up to a wave function renormalization.
We shift the data of the 2-point correlation functions 
for $(\mu^2, \lambda)=(-12.810, 0.6),  (-14.925, 0.7)$
simultaneously in the vertical direction 
by $\alpha_{0.6 \to 0.5}=\log[Z(\lambda=0.5)/Z(\lambda=0.6)]=-0.015(1)$ 
and $\alpha_{0.7 \to 0.5}=-0.056(1)$, respectively, and plot
the shifted data in Fig.\ref{2pt_after_renormalization_oncritical}.
We also shift the data of the 4-point
correlation functions for $(\mu^2, \lambda)=(-12.810, 0.6),  (-14.925, 0.7)$
simultaneously by $2\alpha_{0.6 \to 0.5}$ and $2\alpha_{0.7 \to 0.5}$, respectively, and plot the shifted data in Fig.\ref{4pt_after_renormalization_oncritical}. 
We see a good agreement of both  the shifted 2-point and 4-point correlation functions.
These shifts correspond to a wave function renormalization.
Furthermore, we see that
the above tuned values of $\mu^2$ are consistent with the critical values of $\mu^2$ read off  from Fig.\ref{chi}.
Thus, the agreement of the correlation functions implies that the theories
are universal on the phase boundary as in ordinary field theories.
We do not see the above agreement of the correlation functions in either the UV region with $m=1,2$, or the IR region with $m=14, 15$. We consider the disagreement in the latter region to be caused by an IR cutoff that is introduced when the theory on the fuzzy sphere is mapped to a theory on the plane with infinite volume.

Finally, we examine a connection of the present theory to a CFT. In Fig.\ref{2pt_after_renormalization_oncritical}, we fit seven data points ($m=4,\ldots,10$) of  
$\log \langle \varphi(z_m)\varphi(1)\rangle$ at 
$(\mu^2,\lambda)=(-10.801, 0.5)$
to $-u\log |z-1|+v$ and obtain $u=0.149(2)$ and $v=1.887(1)$. 
This implies that the 2-point correlation function behaves as
\begin{equation}
\langle \varphi(z)\varphi(1)\rangle = \frac{e^v}{|z-1|^u}
\label{2pt function on the boundary}
\end{equation}
for $m=4,\ldots,10$. 
In CFTs, the 2-point correlation function behaves as
\begin{equation}
\left \langle \mathcal{O}(z) \mathcal{O}(z') \right \rangle \sim \frac{1}{|z-z'|^{2\Delta}} \ ,
\end{equation}
where the $\Delta$ is the scaling dimension of the operator $\mathcal{O}(z)$.
Thus, the theory on the phase boundary behaves as a CFT in the UV region.
In the IR region with $11 \leq m \leq 13$, 
our 2-point correlation function deviates universally 
from that in the CFT.
In addition, in a further UV region with $m=3$, it also deviates universally.
These deviations are considered to be an effect of the UV/IR mixing.
It is nontrivial that we observe the behavior of the CFT because field theories on noncommutative spaces are non-local ones.

\section{Conclusion and discussion}
\setcounter{equation}{0}
In this paper, we have studied renormalization in the scalar filed theory
on the fuzzy sphere by Monte Carlo simulation.
We showed that the 2-point and 4-point correlation functions in the disordered
phase are made independent of the UV cutoff up to the wave function renormalization
by tuning the mass parameter or the coupling constant.
This strongly suggests that the theory can be renormalized nonperturbatively in the ordinary sense 
and that the theory is universal up to a parameter fine-tuning.

We also examined the 2-point and 4-point correlation functions on the phase 
boundary beyond which the $Z_2$  symmetry is spontaneously broken.
We found that the 2-point and 4-point correlation functions at different points
on the boundary agree up to the wave function renormalization. This implies that
the critical theory is universal, which is consistent with the above universality
in the disordered phase, because the phase boundary is obtained by a parameter fine-tuning. 
Furthermore, we observed that the 2-point correlation functions behave as those
in a CFT at short distances and deviate universally from those at long distances.
The latter is considered to be due to the UV/IR mixing.

The CFT that we observed at short distances seems to differ from the critical Ising model, because the value of $u$ in \eqref{2pt function on the boundary} disagrees with $2\Delta$, where $\Delta$ is the scaling dimension of the spin operator, $1/8$.
This suggests that the universality classes of the scalar field theory on the fuzzy sphere are totally different from those of an ordinary field theory\footnote{It should be noted that the scaling dimension that we obtained, $\Delta \simeq 0.075 = 3/40$, coincides with that of the spin operator in the tricritical Ising model, which is the $(4, 5)$ unitary minimal model.}.

Indeed, it was reported in \cite{Martin:2004un,Panero:2006bx,Panero:2006cs,
GarciaFlores:2009hf,Das:2007gm} that there exists a novel phase in the theory on the fuzzy sphere that is called the nonuniformly ordered phase \cite{Gubser:2000cd,Ambjorn:2002nj}.
We hope to elucidate the universality classes by studying renormalization
in the whole phase diagram.

\section*{Acknowledgements}
Numerical computation was carried out on the XC40 at YITP at Kyoto University and FX10 at the University of Tokyo.
The work of A.T. is supported in part by a Grant-in-Aid
for Scientific Research
(No. 15K05046)
from JSPS.


\section*{Appendix: Bloch coherent state and Berezin symbol}
\setcounter{equation}{0}
\renewcommand{\theequation}{A.\arabic{equation}}
In this appendix, we summarize the basic properties of  
the Bloch coherent state \cite{Gazeau} and 
the Berezin symbol\cite{Berezin:1974du}.

We use a standard basis $|jm\rangle$ $(m=-j, -j+1,\ldots,j)$
for the spin-$j$ $(=(N-1)/2)$ representation of the $SU(2)$ algebra.
The action of $L_i$ on the basis is given by 
\begin{align}
L_{\pm}|jm\rangle &=\sqrt{(j\mp m)(j \pm m+1)}|j m\pm 1\rangle, \nonumber\\
L_3|jm\rangle &= m |jm\rangle \ ,
\end{align}
where $L_{\pm}=L_1 \pm i L_2$.
The highest-weight state $|jj\rangle$ is considered to correspond to the north pole.
Thus, the state $|\Omega\rangle$ 
that corresponds to a point $\Omega=(\theta,\varphi)$ is obtained
by acting a rotation operator on $|jj\rangle$:
\begin{equation}
|\Omega\rangle=e^{i\theta (\sin\varphi L_1 -\cos\varphi L_2)}|jj\rangle \ .
\label{definition of coherent state}
\end{equation}
(\ref{definition of coherent state}) implies that
\begin{equation}
n_iL_i |\Omega\rangle =j |\Omega\rangle  \ ,
\label{property 1}
\end{equation}
where $\vec{n}=(\sin\theta\cos\varphi,\sin\theta\sin\varphi,\cos\theta)$.
It is easy to show from 
(\ref{property 1}) that 
$|\Omega\rangle$ 
minimizes $\sum_i (\Delta L_i)^2$ with $(\Delta L_i)^2$
being the standard deviation of $L_i$.

It is convenient to introduce the stereographic projection, $z=R\tan \frac{\theta}{2} e^{i\varphi}$.
Then, (\ref{definition of coherent state}) is rewritten as
\begin{equation}
|\Omega\rangle=e^{zL_-/R} e^{-L_3 \log (1+|z/R|^2)} e^{-\bar{z}L_+/R} |jj\rangle \ ,
\label{definition of coherent state 2}
\end{equation}
which gives an explicit form of
$|\Omega\rangle$ as
\begin{equation}
|\Omega\rangle=\sum_{m=-j}^{j}
\left( 
\begin{array}{c}
2j \\
j+m
\end{array}
\right)^{\frac{1}{2}}
\left( \cos \frac{\theta}{2} \right)^{j+m} \left( \sin \frac{\theta}{2} \right)^{j-m} 
e^{i(j-m) \varphi} |jm\rangle \ .
\label{explicit form}
\end{equation}
By using (\ref{explicit form}), one can easily show the following relations:
\begin{align}
& \langle \Omega_1 | \Omega_2 \rangle
=\left( \cos\frac{\theta_1}{2}\cos\frac{\theta_2}{2}+e^{i(\varphi_2-\varphi_1)}
\sin\frac{\theta_1}{2}\sin\frac{\theta_2}{2} \right)^{2j} \ , 
\label{property2}\\
& |\langle \Omega_1 | \Omega_2 \rangle |= \left(\cos \frac{\chi}{2}\right)^{2j} 
\;\;\mbox{with} \;\;
\chi =\arccos (\vec{n}_1\cdot \vec{n}_2) \ , 
\label{property3}\\
& \frac{2j+1}{4\pi} \int d\Omega \ |\Omega\rangle\langle \Omega | =1 \ .
\label{property4}
\end{align}
(\ref{property3}) 
implies that the width of the Bloch coherent state is
$\frac{R}{\sqrt{j}}$ for large $j$.

Denoting the Bloch coherent state $|\Omega\rangle$ by $|z\rangle$,
we rewrite
(\ref{explicit form}) and (\ref{property4}) as
\begin{align}
&|z\rangle = \left(\frac{z/R}{1+|z/R|^2}\right)^j \sum_{m=-j}^j 
\left(
\begin{array}{c}
2j \\
j+m
\end{array}
\right)^{\frac{1}{2}}
\frac{R^m}{z^m} 
|jm\rangle \ , \label{explicit form 2}\\
&\frac{2j+1}{4\pi}4R^2\int \frac{d^2z}{(1+|z/R|^2)^2} \  |z\rangle \langle z| =1 \ ,
\label{property4prm}
\end{align}
respectively.

The Berezin symbol for a matrix $A$ with the matrix size $2j+1$ is defined by
\begin{align}
f_A(\Omega) & = f_A(z,\bar{z}) \nonumber\\
&=\langle \Omega | A | \Omega \rangle \nonumber\\
&=\langle z | A | z \rangle \ .
\end{align}
By using (\ref{explicit form}), it is easy to show that 
\begin{equation}
f_{[L_i,A] }(\Omega)={\cal L}_i f_{A}(\Omega) \ .
\label{derivative}
\end{equation}
(\ref{property4}) implies that
\begin{equation}
\frac{1}{N}\mbox{Tr}(A)=\int \frac{d\Omega}{4\pi} f_A(\Omega)  \ .
\label{trace and integral}
\end{equation}

The definition of the star product for $A$ and $B$ is 
\begin{equation}
f_A\star f_B(\Omega) = f_A\star f_B (z,\bar{z})
=\langle\Omega | AB |\Omega\rangle =\langle z | AB | z \rangle \ .
\end{equation}
Here let us consider a quantity
\begin{equation}
\frac{\langle w | A | z \rangle}{\langle w | z \rangle} \ ,
\end{equation}
which is holomorphic in $z$ and anti-holomorphic in $w$.
Then, one can deform this quantity as follows:
\begin{align}
\frac{\langle w | A | z \rangle}{\langle w | z \rangle}
&=e^{-w\frac{\partial}{\partial z}}
\frac{\langle w | A | z+w \rangle}{\langle w | z+w \rangle} \nonumber\\
&=e^{-w\frac{\partial}{\partial z}}e^{z\frac{\partial}{\partial w}}
\frac{\langle w | A | w \rangle}{\langle w | w \rangle}  \nonumber\\
&=e^{-w\frac{\partial}{\partial z}}e^{z\frac{\partial}{\partial w}}
\langle w | A | w \rangle \nonumber\\
&=e^{-w\frac{\partial}{\partial z}}e^{z\frac{\partial}{\partial w}}
f_A(w,\bar{w})  \ .
\label{holomorphy}
\end{align}
Similarly, one obtains
\begin{equation}
\frac{\langle z | A | w \rangle}{\langle z | w \rangle}
=e^{-\bar{w}\frac{\partial}{\partial \bar{z}}}e^{\bar{z}\frac{\partial}{\partial \bar{w}}}
f_A(w,\bar{w})  \ .
\label{holomorphy2}
\end{equation}
By using (\ref{property4prm}), (\ref{holomorphy}), and (\ref{holomorphy2}),
one can express the star product as
\begin{align}
f_A\star f_B(w,\bar{w})
&=\langle w | AB | w \rangle  \nonumber\\
&=\frac{2j+1}{4\pi}4R^2\int \frac{d^2z}{(1+|z/R|^2)^2}
\frac{\langle w | A | z \rangle}{\langle w | z \rangle}
\frac{\langle z | B | w \rangle}{\langle z | w \rangle}
|\langle w | z \rangle |^2 \nonumber\\
&=\frac{2j+1}{4\pi}4R^2\int \frac{d^2z}{(1+|z/R|^2)^2}
(e^{-w\frac{\partial}{\partial z}}e^{z\frac{\partial}{\partial w}}
f_A(w,\bar{w}) )
(e^{-\bar{w}\frac{\partial}{\partial \bar{z}}}e^{\bar{z}\frac{\partial}{\partial \bar{w}}}
f_B(w,\bar{w}))
|\langle w | z \rangle |^2  \ ,
\label{star product}
\end{align}
which indicates that the star product is noncommutative and non-local.
Furthermore, one can easily show that in the $j\rightarrow\infty$ limit
\begin{equation}
\frac{2j+1}{4\pi}\frac{4R^2}{(1+|z/R|^2)^2}
|\langle w | z \rangle |^2 
\rightarrow \delta^2(z-w) \ .
\end{equation}
This implies that the star product coincides with the ordinary product in the $j\rightarrow\infty$ limit. Namely,
\begin{equation}
f_A\star f_B(w,\bar{w}) \rightarrow f_A(w,\bar{w})f_B(w,\bar{w}) 
\end{equation}
or
\begin{equation}
f_A\star f_B(\Omega) \rightarrow f_A(\Omega)f_B(\Omega) \ .
\label{ordinary product}
\end{equation}

We see from (\ref{derivative}), (\ref{trace and integral}), 
and (\ref{ordinary product}) that
the theory (\ref{action}) reduces to that of (\ref{continuum action}) in the 
$N\rightarrow\infty$ limit at the classical level
if one identifies $f_{\Phi}(\Omega)$ with $\phi(\Omega)$.
However, the authors of 
\cite{Chu:2001xi,Steinacker:2016nsc} showed that
the one-loop effective action in (\ref{action}) differs from that in
(\ref{continuum action}) by finite and non-local terms
since the UV cutoff $N$ is kept finite in calculating
loop corrections.
This phenomenon is sometimes called the UV/IR anomaly.

\end{document}